\documentclass[12pt,preprint]{aastex}

\shorttitle{Subsurface stratification and solar cycle}
\shortauthors{Lefebvre \& Kosovichev}

\begin{document}

\title{Changes in the subsurface stratification of the Sun with the 11-year activity cycle}

\author{S. Lefebvre\altaffilmark{1}}
\affil{Physics and Astronomy Building, UCLA, Los Angeles, CA
90095-1547, USA} \email{lefebvre@astro.ucla.edu}

\and

\author{A. G. Kosovichev\altaffilmark{2}}
\affil{W. W. Hansen Experimental Physics Laboratory, Stanford
University, Stanford, CA 94305-4085, USA}

\begin{abstract}

We report on the changes of the Sun's subsurface stratification
inferred from helioseismology data. Using SOHO/MDI (SOlar and
Heliospheric Observatory/Michelson Doppler Imager) data for the last
9 years and, more precisely, the temporal variation of f-mode
frequencies, we have computed the variation of the radius of
subsurface layers of the Sun by applying helioseismic inversions. We
have found a variability of the ``helioseismic'' radius in antiphase
with the solar activity, with the strongest variations of the
stratification being just below the surface around 0.995$R_{\odot}$.
Besides, the radius of the deeper layers of the Sun, between
0.975$R_{\odot}$ and 0.99$R_{\odot}$ changes in phase with the
11-year cycle.

\end{abstract}


\section{Introduction}

For the last decades, temporal variations in the solar radius has
been a controversial subject. Indeed, measurements made with the
solar astrolabe by F. Laclare in France \citep{Laclare96} and the
Brazilian team \citep{Reis03} showed a variation of the solar radius
in antiphase with the solar activity cycle, while F. No\"el in Chile
\citep{Noel04} using a similar instrument reported a variation in
phase with the solar activity. By using solar continuum intensity
images obtained with the Michelson Doppler Imager (MDI) on board
SOHO, \citet{Emilio00} found a variation of the solar radius of $8.1
\pm 0.9$ mas/yr, but in a recent and more complete study of these
images, \citet{Kuhn04} reported no evidence of solar-cycle visible
radius variations between 1996 and 2004 at any level above 7
milliarcseconds: this is significantly lower than any variation
reported from ground-based observations. Other measurements made by
the Solar Disk Sextant (SDS) experiment onboard strastospheric
ballons \citep{Sofia94,Thuillier05} reported a variation of the
solar radius in antiphase with the solar radius. The solar radius
can be also determined by helioseismic methods \citep{Schou97}.
This, so-called ``seismic'' radius, is related to the subsurface
density stratification and can be compared with the ``photospheric''
radius, as inferred by astrolabe ground-based measurements, for
example, only by using solar models. The ''seismic'' radius probes
the sub-photospheric layers up to a depth of about 15 Mm. So the use
of solar f-mode frequencies to infer the seismic radius is important
for searching for physical changes occuring beneath the photosphere.
Results are, however, conflicting. \citet{Dziem01} reported rates of
the seismic radius change ranging from -3 to 1 km/year during the
rise of cycle 23. Recently, \citet{Dziem05} reexamined the issue by
using SOHO/MDI f modes for 1996-2004, and were unable to obtain
stable results for the seismic radius, and thus they concluded that
the observed variations of the f-mode frequencies should be
explained by the direct effect of magnetic field. However this led
to unrealistically strong random field hidden below the surface. So,
the question whether or not there is a radius variation with the
activity cycle is still debated. Moreover if there are such
variations at the surface, where is their origin? In this letter, we
consider the issue of solar radius and determine solar radius
variations with time for various layers below the surface. The main
difference from the previous investigations is that we do not assume
an uniform change of the radius of subsurface layers, but allow
variations of displacement of these layers with depth. We show that
a stable solution does exist in this case, and find evidence for temporal solar radius variations
in the sub-photospheric layers above 0.97$R_{\odot}$ with an
oscillation, in antiphase with the solar cycle above 0.99$R_{\odot}$
and in phase below. Our results show the localization of these
variations in the upper convective zone. If we extrapolate these results up to the surface, we find a radius change of about 2 km in antiphase with the solar cycle: here, we have to keep in mind that without high-l, we cannot constrain the surface radius better, and that in reality, this variation at the surface can be larger provided it is more localized.

\section{Data}

We used frequencies of solar oscillation modes from 72-day MDI
observing runs, computed by J.
Schou\footnote{\url{http://quake.stanford.edu/$\sim$schou/anavw72z/}}.
We selected only f modes for the observing period 1996-2004. Each
file has a different number of f modes, so we extracted only common
modes from these files and obtained a total of 151 modes ranging
from the angular degree $l=125$ to $l=285$. We have computed the
relative frequency changes $\delta\nu/\nu$ for each degree and
calculated the average over each year and binned every 20 $\mu$Hz.
The reference year has been chosen to be 1996, near the minimum of
the solar cycle\footnote{For this year, data begin from May
$1^{st}$.}. The averaging over 1-year of the data allowed us to
avoid the instrumental 1-year variations. The results are compiled
on Fig. 1. The errorbars are not plotted for clarity of the graph.


The different curves plotted on this figure shows an evidence of
variations in the f-mode frequencies over the solar cycle already
pointed out by \citet{Dziem01}. We emphasize an intriguing
phenomenon at higher $l$, above $\nu=1600$ $\mu$Hz, where the
frequency difference can change of sign, becoming then negative, in
the descending part of the solar cycle. It is particularly puzzling
that the sharp frequency decrease above 1600 $\mu$Hz continues in
the declining phase of the solar cycle whereas the frequencies of
the lower-frequency modes return to their previous solar minimum
values. We don't know how to interpret this change but assume that
it may come from a variation in the near-surface turbulence, which
can affect the frequencies of the f modes confined just below the
surface, in a zone close to the leptocline (from the greek
\textit{leptos}=fine), thin transition layer between the upper
convective zone and the photosphere \citep{Godier01,Rozelot03}. We
used these computed frequency differences (without the binning over
$\nu$) between $l=150$ and 250, which are measured most reliably, as
input parameters in our helioseismic inversion presented in the next
section.

\section{Helioseismic inversion of f modes to infer solar radius variations}

\subsection{Mathematical formalism}

As a starting point for our inversion, we used the equation derived
by \citet{Dziem04} who established a relation between the relative
frequency variations $\delta\nu/\nu$ for f-mode frequencies and the
associated Lagrangian perturbation of the radius $\delta r/r$ of
subsurface layers:
\begin{equation}
\left(\frac{\delta\nu}{\nu}\right)_l=-\frac{3l}{2\omega^2 I}\int
dI\frac{g}{r}\frac{\delta r}{r} \label{eq_radius}
\end{equation}
where $l$ is the degree of the f modes, $I$ is the moment of
inertia, $\omega$ the eigenfrequency and  $g$ the gravity
acceleration.

This equation leads to the asymptotic relation used in the previous
determinations of the solar radius using f modes $\frac{\Delta
\nu_l}{\nu_l}=-\frac{3}{2} \frac{\Delta R_{\odot}}{R_{\odot}}$
\citep{Schou97}, assuming that $\frac{\delta r}{r}$ is constant with
depth. However, we don't make this assumption, and infer
$\frac{\delta r}{r}$ as a function of $r$. Eq. \ref{eq_radius} can
be rewritten as
\begin{equation}
\left(\frac{\delta\nu}{\nu}\right)_l=\int^{R_{\odot}}_{0}
K_l\frac{\delta r}{r} dr \label{eq_radius_2}
\end{equation}
where the kernel $K_l$ is expressed as
\begin{equation}
K_l=-\frac{3l}{2\omega^2_l I_l} \rho \left|\vec{\xi_l}\right|^2 gr
\end{equation}
$\rho$ being the density and $\vec{\xi_l}$ the mode eigenfunction.

To compute each kernel $K_l$, we used the model S
\citep{Christensen96} calibrated to the seismic radius of
\citet{Schou97} ($R_{\odot}=6.9568\times10^5$ km). Fig. 2 shows
three examples of the kernels at $l=$ 150, 200 and 250. The method
used to invert Eq. \ref{eq_radius_2} is the standard Regularised
Least-Square technique \citep{Tikhonov77}, appropriate for this kind
of ill-posed problems. In this method, one has to minimize the
following relation
\begin{eqnarray}
E & = & \sum_{l} \frac{1}{\sigma^2_l}\left(\int^{R_{\odot}}_{0} K_l
y dr - f_l\right)^2 + \alpha \int^{R_{\odot}}_{0} y^2 dr \nonumber
\\ & + & \beta \int^{R_{\odot}}_{0} \left(\frac{dy}{dr}\right)^2 dr
\label{eq_RLS}
\end{eqnarray}
where $y=\frac{\delta r}{r}$, $f_l=\left(\frac{\delta
\nu}{\nu}\right)_l$, $\sigma_l$ the relative incertainty for each
$f_l$, $\alpha$ and $\beta$ the regularisation parameters.


\subsection{Results}

In the inversion process, we restricted the data set using only
modes with $l$ ranging from 150 to 250 measured more accurately than
the other modes. For each year, the minimization of Eq. \ref{eq_RLS}
leads to a solution plotted on Fig. 3. This solution integrated
through Eq. \ref{eq_radius_2} yields the relative frequencies
variations versus the degree $l$ in comparison with the real data.
The reconstructed frequencies fit pretty well the trend of the real
relative frequencies within the errorbars (Fig. 4). Fig. 5 shows the
$\Delta R$ estimated at the surface using
$R_{\odot}=6.9568\times10^5$ km as a reference in the model. Fig. 6
shows the averaging kernels \citep{Ory95} which illustrate the
resolving power of these inversions. It is quite clear that the
spatial resolution (FWHM) is about $0.005R_\odot$ in the region
between $0.98R_\odot$ and $0.998R_\odot$. Like in other inverse
problems, this means that the variations of the smaller scale and
outside this region cannot be resolved without additional
constraints. In Fig. 7 we give three examples of test inversions
with artificial f-mode frequency data (applying the observed error
estimates), which illustrate the accuracy and limitations of these
inversions.


The main characteristics of our solution are:
\begin{enumerate}
    \item Fig. 3 shows no significant changes in the variation
    of the subsurface layers depth below $0.97R_{\odot}$.
    In this layers, our inversions loose the spatial resolution
    as follows from Fig. 6. So, with the currently available data,
    it is not possible to measure localized variations below
    $0.97R_{\odot}$. However, if there were an uniform change
    of the radius in subsurface layers (including $0.97R_\odot$ and
    below) of the order of 1-3 km as discussed by \citet{Dziem01},
    it could be detected by the inversion procedure as illustrated
    in the top panel of Fig.7 (no assumption was made about the
    functional form of the solution).
    \item Fig. 3 shows non-monotonic changes in the stratification
    with the inner layer (below $0.99R_{\odot}$) moving up during
    the increase of activity (compression) and the outer layer
    (above $0.99R_{\odot}$) moving down (relaxation).
    The precise localization of these layers is uncertain
    because have a characteristic width of about $0.005R_{\odot}$. The test
    inversion in Fig.7 (middle panel) shows that this uncertainty
    can be about the half-width of the averaging kernels, about
    $0.003R_\odot$.
    \item Fig. 5 estimates that the near-surface variations are
    in antiphase with the solar cycle with an amplitude of the order of 2 km.
    However, the sensitivity of our inversions is quite low at the surface,
    and localized variations of the surface radius may not be detected (see
    a test inversion in bottom panel of Fig.7). High-degree f-mode
    data are required to improved the surface estimates.
\end{enumerate}

Note that if we used all the modes available (i.e. 151 modes), we
cannot find a stable solution that fits the last part of the curves
above $\nu=1600$ $\mu$Hz (see Fig. 1). We suppose that the behavior
of the curves in this range of frequencies could be due to the
influence of turbulence and magnetic fields very near the surface.

\section{Discussion}

First of all, Fig. 3 shows a temporal variation of the solar radius
in the subsurface layers during the solar cycle. This oscillation is
centered around $r=0.99R_{\odot}$ with a width of about
$0.03R_{\odot}$. This anomaly is composed of two parts: the first,
below $r=0.99R_{\odot}$, is in phase with the solar cycle and has a
maximal amplitude of about 10 km; the second part above
$r=0.99R_{\odot}$, is in antiphase with the solar cycle and reaches
larger amplitudes, of about 26 km. These variations indicate the
presence of a changing with the solar cycle of a physical structure
that could be described as a very thin transition layer, siege of
the variation of the solar radius. This transition layer is located
here approximately at $0.99R_{\odot}$ and it is linked to changes in
the upper convective zone caused by magnetic fields. We estimate a
variation of the seismic radius at the surface of about 2 km at the
maximum of the cycle 23. However, the surface radius is poorly
constraint with the currently available set of medium-l f-mode
frequencies.

Our helioseismic inversion computations have put in evidence a
variability of the solar radius in the subsurface layers which are
extended to the surface. The results are generally consistent with
previous conclusions that solar-cycle variations in the solar radius
are confined to the outermost layers of the Sun
\citep{Antia04,Dziem05}. This variability is localized in a
double-structure layer centered at $0.99 R_{\odot}$: in the deeper
part, between $0.97 R_{\odot}$ and $0.99 R_{\odot}$, the radius
varies in phase with the solar cycle, whereas this is opposite in
the upper part above $0.99 R_{\odot}$, where the variability become
in antiphase.  Thus we confirm the fact that the surface layers of
the Sun are shrinking during the ascending phase of the solar cycle
and is relaxing after the maximum. However, these changes are not
uniform with depth. In a near future, it would be interesting to
inspect more in details the behavior very close to the surface by
looking at changes of the higher degree modes, above $l=250$, where
a second thin layer may take place.

As a conclusion, we would like to emphasize our most significant result: the change in radius goes from being in phase with the solar cycle in the deeper layers to out of phase in the shallower layers with a transition at $0.99 R_{\odot}$. This result could eventually lead to a deeper understanding of the physics behind the changes.

\acknowledgments

This work utilizes data from SOHO/MDI and we thank J. Schou for providing the
frequencies files.

\clearpage

\begin{figure}
\includegraphics[angle=-90,width=15cm]{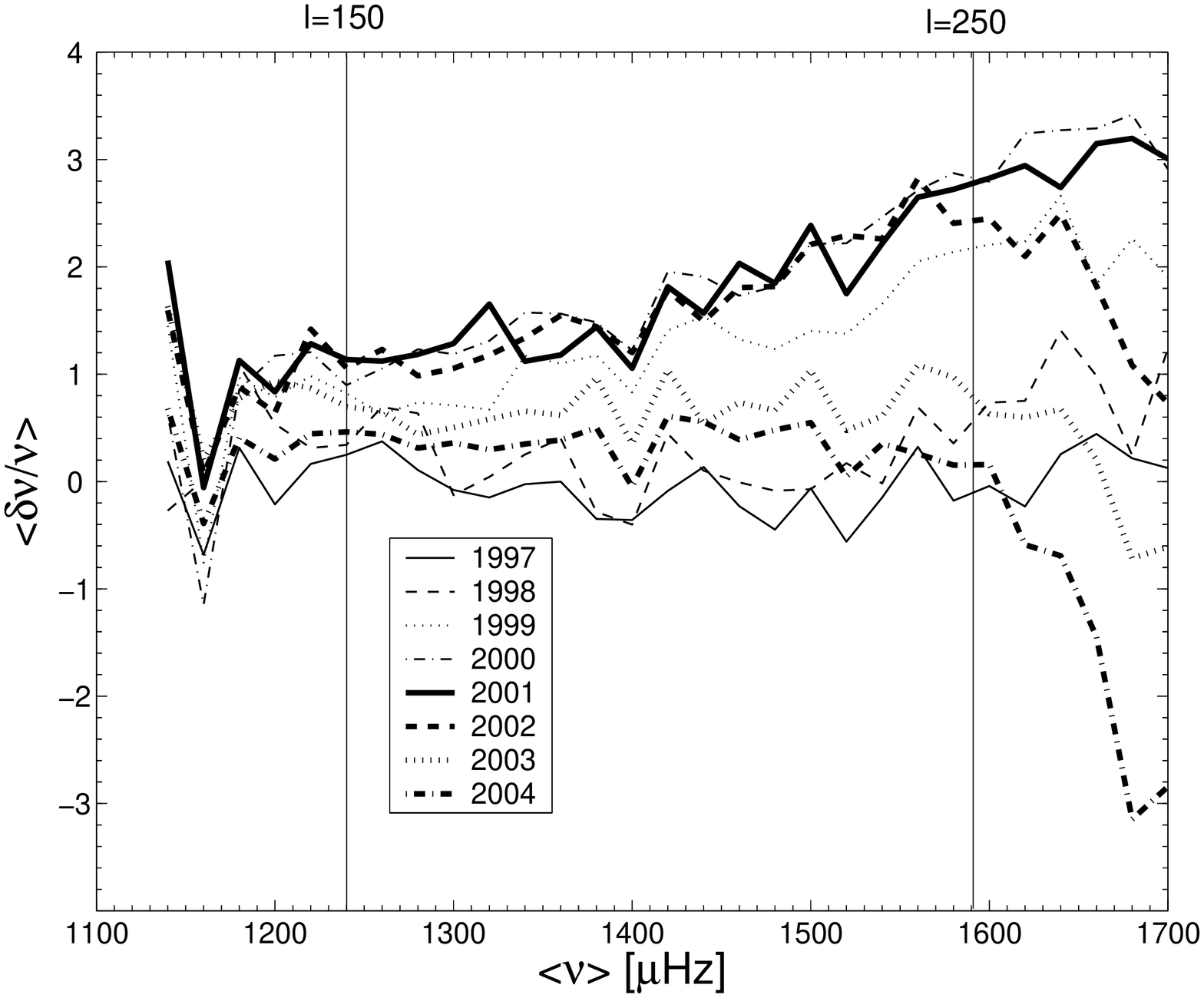}
\caption[f1.eps]{Average relative frequency differences in f-mode
$\left\langle \delta\nu/\nu \right\rangle$ as a function of
$\left\langle \nu \right\rangle$, average frequencies binned every
20 $\mu$Hz. The reference year is 1996 and the errorbars have not be
plotted for clarity of the graph. The f modes chosen for our study
have frequencies between the limits represented by vertical lines.}
\label{figure1}
\end{figure}

\clearpage

\begin{figure}
\includegraphics[angle=-90,width=15cm]{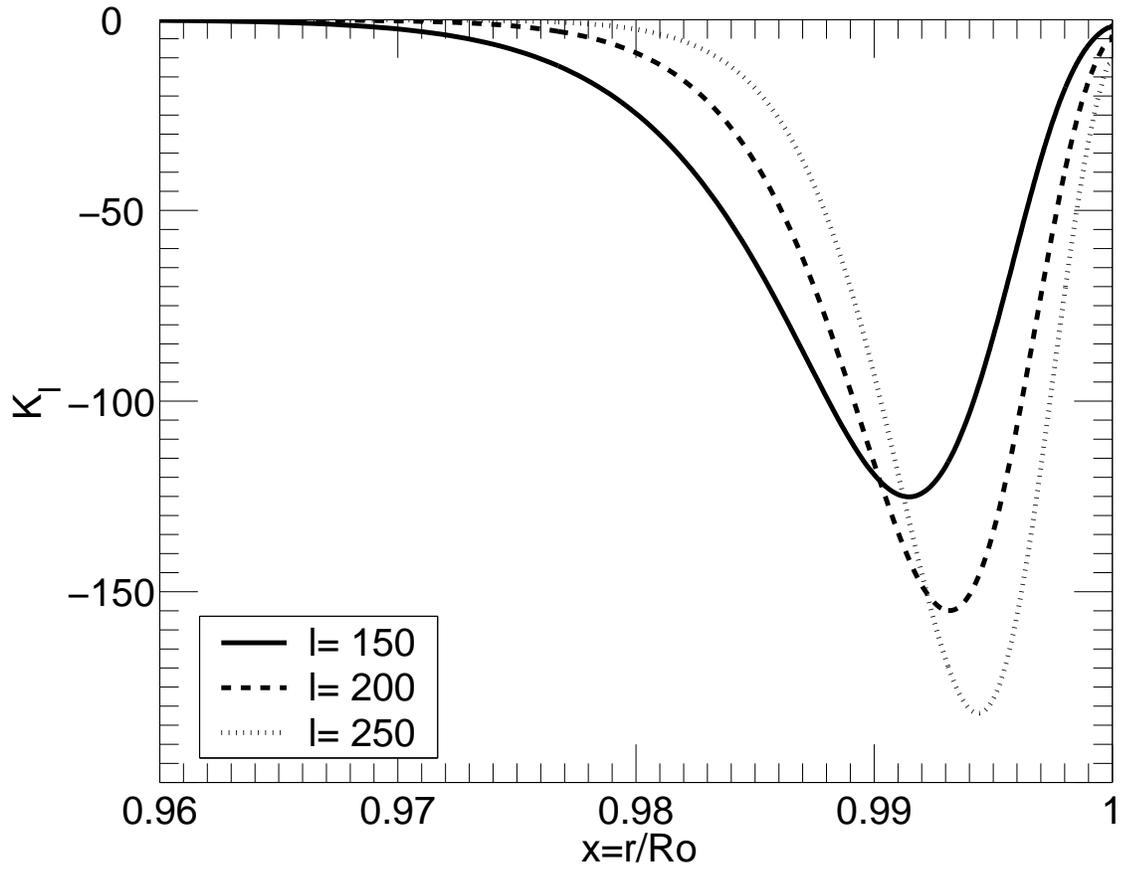}
\caption[f2.eps]{Example of kernels $K_l$ at $l=$ 150, 200 and 250.}
\label{figure2}
\end{figure}

\clearpage

\begin{figure}
\includegraphics[angle=-90,width=15cm]{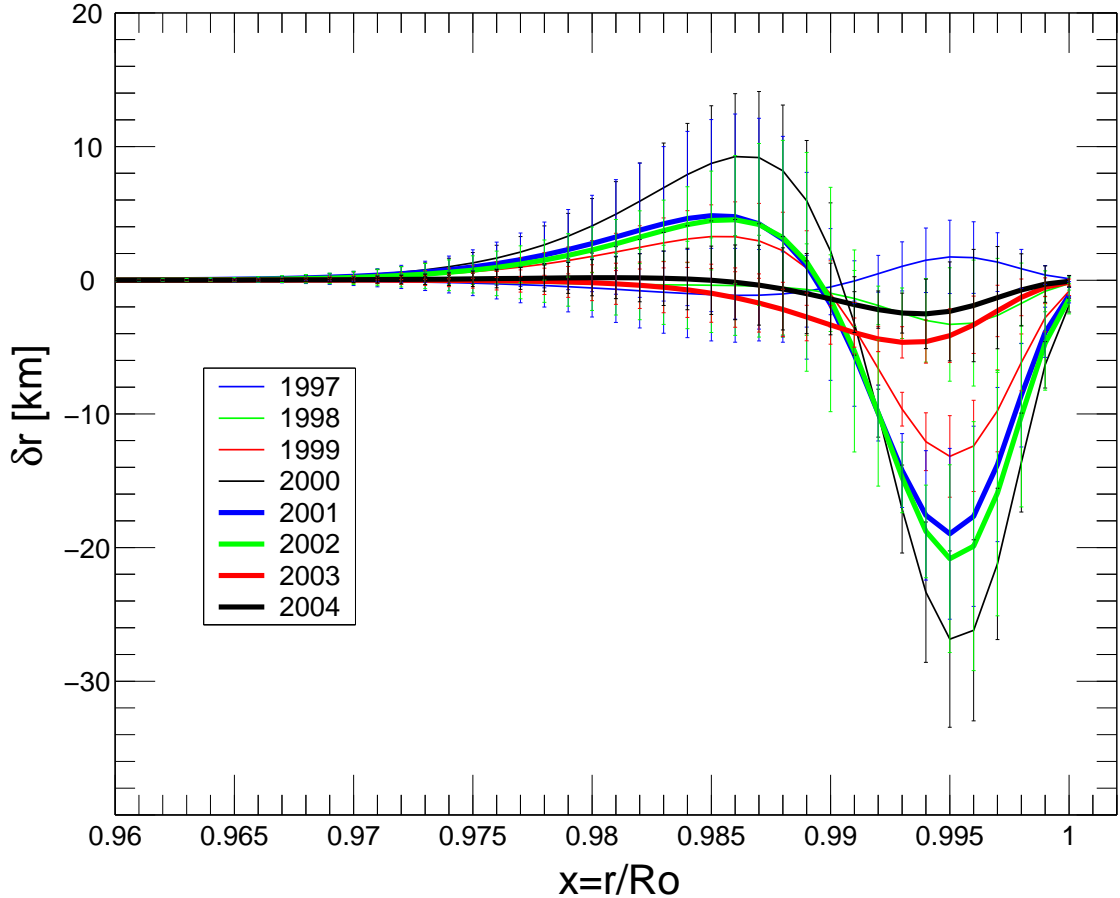}
\caption[f3.eps]{$\left\langle \delta r \right\rangle$ as a function
of the fractional radius $x=r/R_{\odot}$, obtained as a solution of
the minimization of Eq. \ref{eq_RLS}. Notice the behavior of the
curve near $x=0.99$. The errors are the standard deviation after
average over a set of random noise added to the relative frequencies
reconstructed in Fig. 4.} \label{figure3}
\end{figure}

\clearpage

\begin{figure}
\includegraphics[angle=-90,width=15cm]{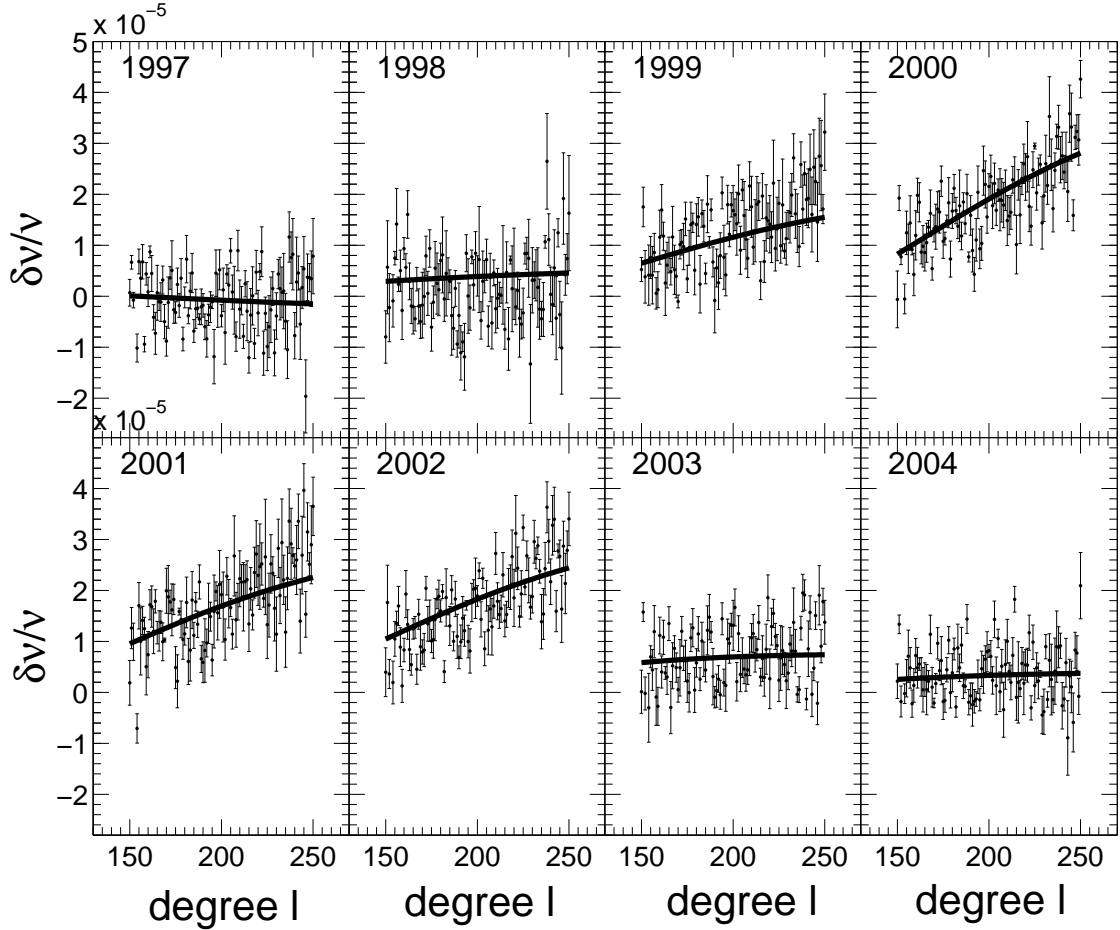}
\caption[f4.eps]{For each year, $\left\langle \delta\nu/\nu
\right\rangle$ as a function of the degree $l$. The reference year
is 1996 and the errors are the relative incertainties. The solid
curve is the results of direct integration of Eq. \ref{eq_radius_2}
providing the solution of Fig. 3.} \label{figure4}
\end{figure}

\clearpage

\begin{figure}
\includegraphics[angle=-90,width=15cm]{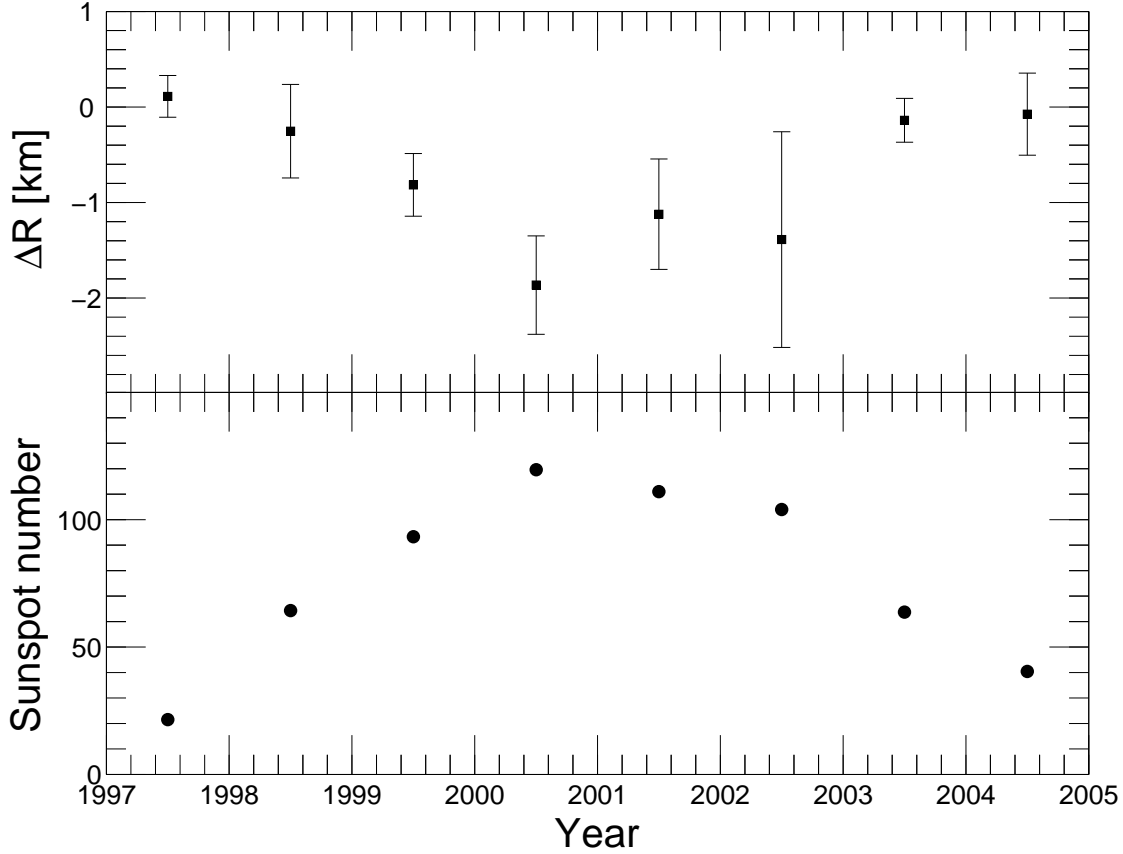}
\caption[f5.eps]{Top: temporal variation of $\Delta R$ near the
solar surface at $r=R_{\odot}$ from the solution of Fig. 3. Bottom:
Variation of the sunspot number for the same period. The variation
of the seismic radius at the surface is found in antiphase with the
solar cycle with an amplitude of about 2 km. It is important to keep in mind that without high-l, we cannot constrain the surface radius better, and that in reality, this variation at the surface can be larger provided it is more localized.} \label{figure5}
\end{figure}

\clearpage

\begin{figure}
\includegraphics[angle=-90,width=15cm]{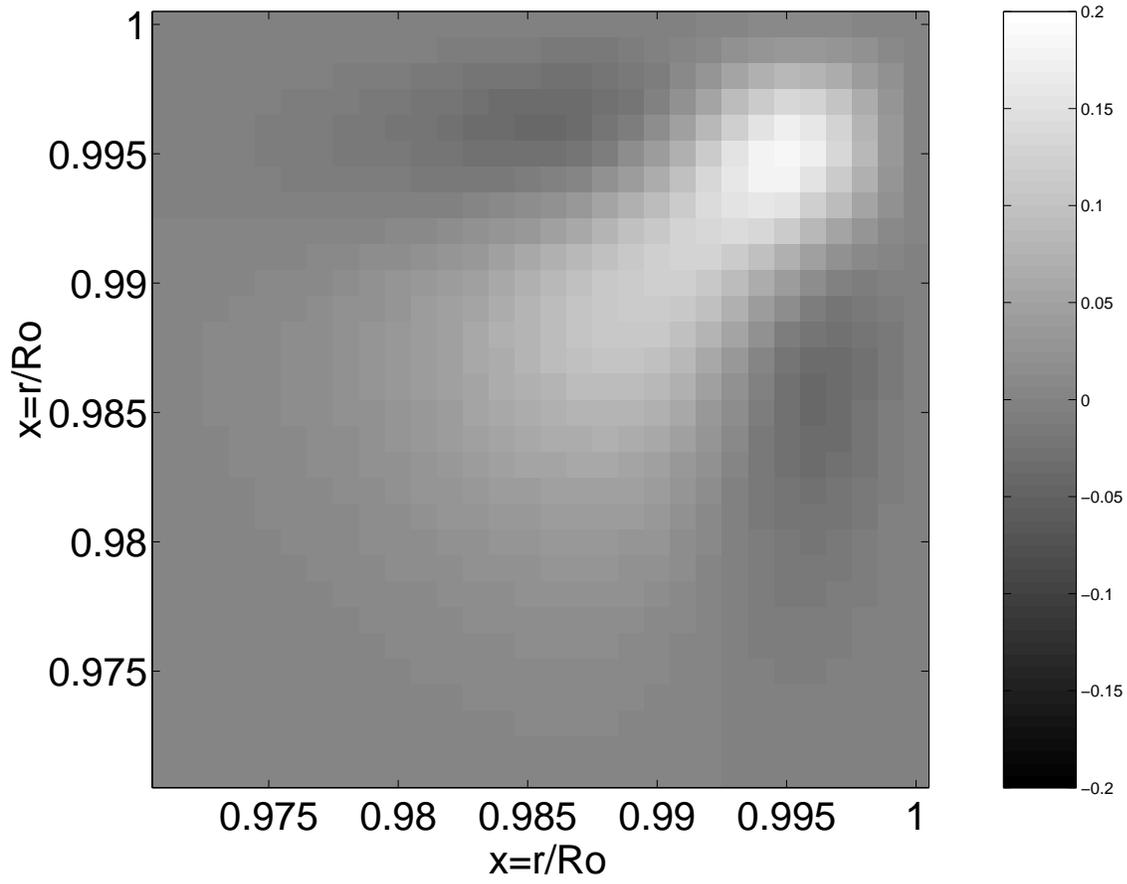}
\caption[f6.eps]{Amplitude of the averaging kernels versus the
fractional radius x.} \label{figure6}
\end{figure}

\clearpage

\begin{figure}
\includegraphics[angle=-90,width=15cm]{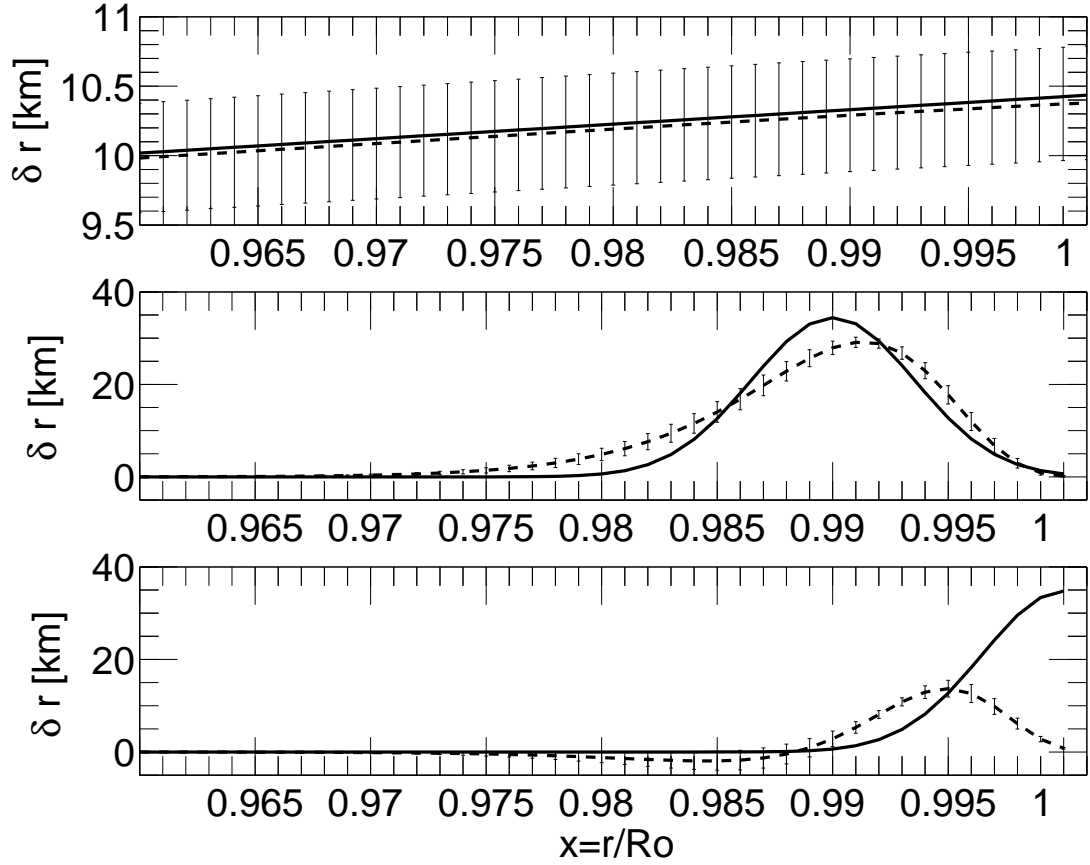}
\caption[f7.eps]{Inversion tests with artificial data; from top to
bottom: the initial $\delta r/r$ is a constant, a gaussian with a
width of 0.005 and centered on x=0.99, a gaussian with a width of
0.005 and centered on x=1. In all cases, the solid line is the
initial data and the dashed curve is the result after inversion.}
\label{figure7}
\end{figure}


\clearpage

\begin{thebibliography}{}

\bibitem[Antia \& Basu(2004)]{Antia04}
Antia, H. M., \& Basu, S. 2004, in ESA-SP 559, Proceedings of the
SOHO 14 / GONG 2004 Workshop: Helio- and Asteroseismology: Towards a
Golden Future, ed. D. Danesy, 301
\bibitem[Christensen-Dalsgaard et al.(1996)]{Christensen96}
Christensen-Dalsgaard et al. 1996, Science, 272, 1286
\bibitem[Dziembowski et al.(2001)]{Dziem01}
Dziembowski, W. A. et al., 2001, \apj, 553, 897
\bibitem[Dziembowski \& Goode(2004)]{Dziem04}
Dziembowski, W. A., \& Goode, P. R. 2004, \apj, 600, 464
\bibitem[Dziembowski \& Goode(2005)]{Dziem05}
Dziembowski, W. A., \& Goode, P. R. 2005, \apj, 625, 548
\bibitem[Emilio et al.(2000)]{Emilio00}
Emilio, M. et al., 2000, \apj, 543, 1007
\bibitem[Godier \& Rozelot(2001)]{Godier01}
Godier, S., \& Rozelot, J. P. 2001, \solphys, 199, 217
\bibitem[Kuhn et al.(2004)]{Kuhn04}
Kuhn, J. R. et al., 2004, \apj, 613, 1241
\bibitem[Laclare et al.(1996)]{Laclare96}
Laclare, F. et al., 1996, \solphys, 166, 211
\bibitem[No\"el(2004)]{Noel04}
No\"el, F. 2004, \aap, 413, 725
\bibitem[Ory \& Pratt(1995)]{Ory95}
Ory, J., \& Pratt, R. G. 1995, Inverse Problems, 11, 397
\bibitem[Reis Neto et al.(2003)]{Reis03}
Reis Neto, E. et al., 2003, \solphys, 212(1), 7
\bibitem[Rozelot \& Lefebvre(2003)]{Rozelot03}
Rozelot, J. P., \& Lefebvre, S. 2003, Lecture Notes in Physics, 599,
4
\bibitem[Schou et al.(1997)]{Schou97}
Schou, J. et al., 1997, \apj, 489, L197
\bibitem[Sofia et al.(1994)]{Sofia94}
Sofia, S. et al., 1994, \apj, 427, 1048
\bibitem[Thuillier et al.(2005)]{Thuillier05}
Thuillier, G. et al., 2005, Advances in Space
Research, 35, 329
\bibitem[Tikhonov \& Arsenin(1977)]{Tikhonov77}
Tikhonov, A. N., \& Arsenin, V. Y. 1977, Solutions of ill-posed
problems, Washington D. C.: Winston
\end{thebibliography}
\end{document}